\documentclass[nofootinbib,onecolumn,superscriptaddress]{revtex4}

\usepackage{graphicx,rotating}
\usepackage{hyperref}
\usepackage[english]{babel}
\usepackage{amsmath}
\usepackage{amssymb}
\usepackage{slashed}

\newcommand{\T}[1]{#1^T} 
\newcommand{\hc}[2][]{#2^{\dagger #1}} 
\newcommand{\abs}[1]{|#1|} 
\newcommand{\cd}{\mathcal{D}} 

\newcommand{\mdm}{M_{\text{DM}}}
\newcommand{\mg}{M_{\text{GUT}}}

\newcommand{\tabref}[1]{Table~\ref{#1}} 
\newcommand{\order}[1]{{\mathcal O}(#1)}

\newcommand{\vev}[1]{\langle#1\rangle}

\newcommand{\ie}{{\it i.e.}}

\newcommand{\be}{\begin{equation}}
\newcommand{\ee}{\end{equation}}
\newcommand{\br}{\begin{eqnarray}}
\newcommand{\bea}{\begin{eqnarray}}
\newcommand{\eea}{\end{eqnarray}}
\newcommand{\er}{\end{eqnarray}}
\newcommand{\ba}{\begin{array}}
\newcommand{\ea}{\end{array}}
\newcommand{\bi}{\begin{itemize}}
\newcommand{\ei}{\end{itemize}}
\newcommand{\bn}{\begin{enumerate}}
\newcommand{\en}{\end{enumerate}}
\newcommand{\bc}{\begin{center}}
\newcommand{\ec}{\end{center}}

\newcommand{\gsim}{\lower.7ex\hbox{$\;\stackrel{\textstyle>}{\sim}\;$}}
\newcommand{\lsim}{\lower.7ex\hbox{$\;\stackrel{\textstyle<}{\sim}\;$}}

\begin{document}


\title{\bf Symmetryless Dark Matter }

\author{Yuji Kajiyama}
\affiliation{National Institute of Chemical Physics and Biophysics, 
Ravala 10, Tallinn 10143, Estonia} 
\affiliation{Akita Highschool, Tegata-Nakadai 1, Akita, 010-0851, Japan}
\author{ Kristjan Kannike}
\affiliation{National Institute of Chemical Physics and Biophysics, 
Ravala 10, Tallinn 10143, Estonia} 
\affiliation{Scuola Normale Superiore and INFN, Piazza dei Cavalieri 7, 56126 Pisa, Italy}
\author{ Martti Raidal}
\affiliation{National Institute of Chemical Physics and Biophysics, 
Ravala 10, Tallinn 10143, Estonia}


\begin{abstract}
It is appealing to stabilize dark matter by the same discrete symmetry that is used to explain the structure of quark and lepton mass matrices. 
However, to generate the observed fermion mixing patterns, any flavor symmetry must necessarily be broken, rendering dark matter unstable. 
 We study singlet, doublet and triplet $SU(2)$ multiplets of both scalar and fermion dark matter candidates and
enumerate the conditions under which no $d < 6$ dark matter decay operators are generated even in the case if the flavor symmetry
is broken to nothing. We show that
the VEVs of flavon scalars transforming as higher multiplets ({\it e.g.} triplets) of the flavor group must be at the electroweak scale. 
The most economical way for that is to use SM Higgs boson(s) as flavons. Such models can be tested by the LHC experiments. 
This scenario requires the existence of additional Froggatt-Nielsen scalars that generate hierarchies in Yukawa couplings.
We study the conditions under which large and small flavor breaking parameters can coexist without destabilizing the dark matter.
\end{abstract}
 
 \maketitle

\section{Introduction}
\label{sec:introduction}

The existence of cold dark matter (DM) is currently the firmest evidence for physics beyond the standard model~\cite{Komatsu:2010fb}. 
The easiest way to stabilize weak scale dark matter particle lifetime over cosmological scales is to impose an exact $Z_{2}$ parity~\cite{Krauss:1988zc,Kadastik:2009dj}. Other Abelian symmetries like $Z_{3}$ have also been studied in this context~\cite{Ma:2007gq,DEramo:2011ib,Agashe:2010tu,Agashe:2010gt,Batell:2010bp}. However, there is no reason why the stabilizing symmetry should be Abelian. Indeed, the study of discrete non-Abelian symmetries is well motivated because these are believed to play a fundamental role in flavor physics. The observed hierarchy of masses and the pattern of mixing angles, in particular the large mixing angles in the lepton sector, remains an enigma. In recent years there has been some progress in explaining 
the possible tribimaximal mixing~\cite{Harrison:2002er} in the neutrino sector with help of discrete flavor groups like $A_{4}$ or $S_{4}$. (See \cite{Ishimori:2010au,Grimus:2011fk} for extensive reviews of many discrete flavor groups and \cite{Altarelli:2010gt} for a review of discrete flavor models.) 
Is it possible to address the two questions, the stability of dark matter and the pattern of flavor mixing, in the same framework of
 non-Abelian discrete symmetries?

 Besides economy, this approach can have benefits like allowing dark matter to decay only into leptons, not quarks. Such a dark matter with lifetime of about $\tau = 10^{26}$~s \cite{Arvanitaki:2008hq} can explain the results of the PAMELA \cite{Adriani:2008zr,Adriani:2008zq} and ATIC \cite{:2008zzr} cosmic ray experiments. While it is also possible that the cosmic ray excesses seen in these experiments are of astrophysical origin, they put a lower bound on dark matter decay rate\footnote{The WMAP bound on lifetime, $\tau > 123 \times 10^{9}$~years or $4 \times 10^{18}$~s \cite{Ichiki:2004vi}, is far more lenient.}.
 Because of its physical appeal, several papers have addressed this issue. 
As shown in \cite{Haba:2010ag}, it is impossible to explain the PAMELA results with an Abelian $Z_{N}$ symmetry while it becomes possible with a non-Abelian flavor symmetry.
 Dark matter stabilized by an unbroken $Z_{2}$ subgroup of flavor symmetry in the neutrino sector is considered in \cite{Hirsch:2010ru,Boucenna:2011tj,Toorop:2011ad}. Recent papers consider dark matter stabilized by $D_{3}$ \cite{Adulpravitchai:2011ei}, $D_{4}$ \cite{Meloni:2011cc} and $D_{6}$ \cite{Kajiyama:2011fe}. Decaying non-Abelian dark matter has also been discussed in \cite{Esteves:2010sh, Daikoku:2010ew,Kajiyama:2010sb}.

The standard scenario in flavor model building is to put the Standard Model (SM) fields into multiplets of the flavor group $G_{\text{f}}$ and introduce new scalar fields -- flavons -- whose vacuum expectation values (VEVs) break $G_{\text{f}}$ in such a way as to give correct fermion mixing angles. 
In the end, the non-Abelian part of the flavor group $G_{\text{f}}$ is completely broken. In addition, usually one or more $Z_{N}$ symmetries are put in by hand to keep, for example, charged lepton flavons from interaction with neutrinos and vice versa. One of these $Z_{N}$ could be used to stabilize dark matter. Contrary to this procedure, 
in this paper we take the position that imposition of additional $Z_{N}$ symmetries by hand is not well justified and it is preferable to get discrete non-Abelian groups from breaking of some non-Abelian gauge group like $SO(3)$ or $SU(3)$ \cite{Frampton:1999hk,deMedeirosVarzielas:2005qg,Hagedorn:2006ug,Luhn:2008sa,Berger:2009tt,Adulpravitchai:2009kd,Grimus:2010ak,Luhn:2011ip}. 

The non-Abelian flavor symmetries, however, do not explain the hierarchy of fermion masses. To address the latter one invokes an additional Froggatt-Nielsen (FN) mechanism \cite{Froggatt:1978nt}. One introduces another scalar which is, as a rule, singlet under the non-Abelian part of the gauge group but charged under a $U(1)$ or $Z_{N}$. The powers of the ratio of the VEV of the Froggatt-Nielsen scalar to the flavor breaking scale give the mass hierarchy.
Flavons are usually chosen to be singlets under the SM gauge group to avoid phenomenological complications. The scale of flavor symmetry breaking is taken to the Grand Unified (GUT) scale or even to the Planck scale. If flavons are SM singlets, they naturally have masses and VEVs at the same scale. These models, while appealing, cannot be tested in a direct way.

It is more economical to consider breaking the electroweak symmetry and the flavor symmetry at the same time by Higgs fields that form a
 multiplet of the flavor group $G_{\text{f}}$. This possibility has  been explored in the context of various flavor groups such as $Q_4$ \cite{Aranda:2011dx}, $Q_6$ \cite{Babu:2009nn,Babu:2011mv,Kaburaki:2010xc,Kifune:2007fj}, $A_4$ \cite{Altarelli:2008bg,Bazzocchi:2008rz,Chao:2007ms,Chen:2005jm,Ciafaloni:2009ub,Frampton:2008ci,Fukuyama:2010ff,Grimus:2008tm,Haba:2010ag,He:2006dk,He:2006qd,Hirsch:2008mg,Hirsch:2008rp,Lavoura:2007dw,Ma:2001dn,Ma:2002yp,Ma:2004zd,Ma:2004zv,Ma:2005mw,Ma:2005qf,Ma:2006sk,Ma:2006wm,Ma:2008ym,Ma:2009wi,Ma:2010gs,Ma:2011yi,Machado:2007ng,Machado:2010uc,Machado:2011gn,Mitra:2009jj,Mondragon:2007nk,Morisi:2007ft,Morisi:2009sc,Morisi:2011pt,Toorop:2010ex,Toorop:2010kt,Zee:2005ut}, $T'$ \cite{Aranda:2010im,BenTov:2011zv,Eby:2008uc,Eby:2009ii,Eby:2011ph,Frampton:2007et,Frampton:2008bz,Frampton:2008vf,Frampton:2009fw,Frampton:2010uw}, $S_3$ \cite{Araki:2005ec,Grimus:2006wy,Kaneko:2006wi,Kaneko:2007ea,Kobayashi:2003fh,Koide:2005ep,Koide:2005za,Koide:2006dn,Kubo:2003iw,Kubo:2004ps,Ma:1999hh,Ma:2004zd,Mitra:2008bn,Mondragon:2007af,Mondragon:2007nk,Morisi:2006pf,Morisi:2011pm}, $S_4$ \cite{Cai:2006mf,Daikoku:2011mq,Dong:2010zu,Hagedorn:2006ug,Lee:1994qx,Ma:2005pd,Morisi:2010rk,Parida:2008pu,Parida:2010jj,Patel:2010hr,Zhang:2006fv}, $T_7$ \cite{Cao:2010mp,Cao:2011cp}, $T_{13}$ \cite{Hartmann:2011pq}, $\Delta(27)$ \cite{Grimus:2008tt,Haba:2005ds,Howl:2009ds,Ma:2006ip,Ma:2007wu} and $\Delta(54)$ \cite{Escobar:2011mq}. Such models face more stringent constraints because they can induce flavor changing neutral currents, lepton number violating decays and other signals ~\cite{Bhattacharyya:2010hp,Lam:2010xm,Toorop:2010ex,Toorop:2010kt,Cao:2011df}, but these constraints and the possibility to discover the new scalars directly at the LHC makes them testable.

In this paper we study, in a systematic way, the conditions under which the existing dark matter of Universe could result from non-Abelian 
discrete flavor symmetries if the latter are completely broken and no additional/remnant $Z_N$ symmetries exist.
Assuming that the dark matter consists of weakly interacting massive particles (WIMP), we classify all decay operators operators of $SU(2)$ singlet, doublet and triplet scalar and fermion dark matter candidates. (Higher $SU(2)$ multiplet dark matter has been considered before \cite{Deshpande:1977rw,Ma:2006km,Barbieri:2006dq,LopezHonorez:2006gr} or higher multiplets \cite{Hambye:2009pw,Cirelli:2005uq,Cirelli:2007xd,Cirelli:2008id,Cirelli:2008jk,Cirelli:2009uv,DelNobile:2009st}, but not in the context of discrete flavor groups.) We show, in a model independent way, that dark matter with an acceptable lifetime can arise from non-Abelian flavor symmetries 
if the symmetry forbids all dark matter decay operators up to $d < 6$ and the VEVs of the non-Abelian flavons are at the elecroweak scale. (In the context of PAMELA, also the $d = 6$ decay operators involving quarks may have to be suppressed.) This is is the most interesting flavor physics case because the flavons may be accessible in the LHC experiments and such a scenario may be directly testable if the fundamental scalars will be discovered at the LHC. 

To make this scenario realistic, however, the flavor models must necessarily explain the fermion mass hierarchies, \ie, contain additional flavons with large VEVs. Therefore we study under which conditions those large VEVs are compatible
with the non-Abelian origin of flavor mixings without inducing too fast dark matter decays. 
We find that such a situation may, indeed, occur even if the flavor group is entirely broken. 
The requirement this to happen is that a partial remnant $Z_N \subset G_\text{f}$ survives in a particular fermion subsector of the model and
 the dark matter transforms nontrivially under this partial remnant, while the symmetry is broken if the entire model is considered. 
For example, the neutrino sector flavons can have large VEVs if dark matter is odd under the remnant $Z_2$ that is unbroken in this sector. 
The mass hierarchy of fermions has to arise from a separate Froggatt-Nielsen flavons that must be singlets (as usual) under the non-Abelian flavor group.
We exemplify those findings by extending the flavor model \cite{Cao:2010mp} based on $T_7$ symmetry.

The structure of this paper is the following. In Section~\ref{sec:dm:decay:ops} we calculate and present dark matter decay operators for both scalars and fermions that are $SU(2)$ singlets, doublets or triplets and consider bounds on dark matter decay rates and the dimension of decay operators. 
In Section~\ref{sec:breaking:to:nothing} we discuss suppression of the dark matter decay operators that originate from breaking the flavor group to nothing. 
We systematize our findings in this Section. In Section~\ref{sec:ex} we present an example model based on $T_7$ that illustrates our general results.
We conclude in Section~\ref{sec:conclusions}.

\section{Decay Operators and Lifetime of Dark Matter}
\label{sec:dm:decay:ops}

We list the decay operators for the $SU(2)$ scalar singlet $S$, doublet $\Phi$ and triplet $\Delta$, and fermion singlet $N$, doublet $L$ and triplet $\Sigma$ dark matter in \tabref{tab:dm:decay:ops}. Neutral $SU(2)$ singlets have hypercharge $Y_{S,N} = 0$. $SU(2)$ doublets with a neutral component have $\abs{Y} = 1$. For definiteness we have $Y_{\Phi} = 1$ as for the SM Higgs doublet and $Y_L = -1$ as for the SM lepton doublets (likewise, our $L$ is left-handed). For $SU(2)$ triplets that contain a neutral component there are two possible hypercharges, $\abs{Y} = 0, 2$. We consider $Y_{\Delta} = 0, 2$ and $Y_{\Sigma} = 0, -2$ for definiteness (the nonzero values are chosen in analogy to the hypercharges of the triplet in type II seesaw \cite{magg:1980,lazarides:1981,mohapatra:1981,cheng:1980} and the right-handed electron, respectively). Our fermion triplets are right-handed. 

Neutral singlet fermions can have Majorana mass, but doublets and triplets need Dirac partners with opposite chirality and hypercharge to get mass. However, the decay operators of the Dirac partner to some fermion $\psi$ are obtained from the given operators with the substitution $\psi \to \mathcal{C} \psi^\dagger$, where $\mathcal{C}$ is the charge conjugation matrix.

All scalar singlet decay operators are derived by multiplying SM singlet operators by the singlet scalar field $S$. There are no $d = 4$ decay operators for $S$ because there are no $d = 3$ SM singlet operators. The $d = 5$ operators come from multiplying SM Lagrangian terms by $S$. The single $d = 6$ operator comes from multiplying the unique $d = 5$ SM operator by $S$.
The decay operators for the scalar doublet $\Phi$, fermion singlet $N$ and fermion doublet $L$ are obtained from SM operators given in \cite{Buchmuller:1985jz,Grzadkowski:2010es} by substituting one dark matter field for the respective SM field. We use the notation $\tilde{H} \equiv i \sigma_2 H^{*}$ and $\hc{\phi} \overleftrightarrow{\cd}_{\mu} \phi \equiv i \hc{\phi} (D_{\mu} - \overleftarrow{D}_{\mu}) \phi$ and $\hc{\phi} \overleftrightarrow{\cd}_{\mu}^{I} \phi \equiv i \hc{\phi} (\tau^{I} D_{\mu} - \overleftarrow{D}_{\mu} \tau^{I}) \phi$.

The fermion singlet operators are consistent with those previously published in \cite{Haba:2010ag}. 

 It is notable that there is only small number of $d < 6$ decay operators for all scalar and fermion dark matter candidates. Applying our idea of dark matter without symmetry requires that only very small number of decay operators, in most cases two, are  absent. The $d =6$ operators are listed for completeness. We have not checked that they form a complete basis, \ie, those operators may be linearly dependent on each other, but they are general in the context of discrete flavor.

\begin{table}[hp]
\caption{DM decay operators.} 
\centering
\begin{tabular}{cp{0.8\textwidth}}
Dimension & DM decay operators \\
\hline
\multicolumn{2}{c}{Scalar Singlet $S$ DM decay} \\
3 & $\abs{H}^{2} S$ \\
4 & -- \\
5 & $\bar{\ell} H e S$, $\bar{q} H d S$, $\bar{q} \tilde{H} u S$, $\abs{H}^{4} S$, $\abs{\cd_{\mu} H}^{2} S $, $\bar{\ell} \slashed{\cd} \ell S$, $\bar{e} \slashed{\cd} e S$, $\bar{q} \slashed{\cd} q S$, $\bar{d} \slashed{\cd} d S$, $\bar{u} \slashed{\cd} u S$ \\
6 & $(\hc{\tilde{H}} \ell)^{T} \mathcal{C} (\hc{\tilde{H}} \ell) S$ \\
\multicolumn{2}{c}{Scalar doublet $\Phi$ DM decay} \\
4 & $\bar{\ell} \Phi e$, $\bar{q} \Phi d$, $\bar{q} \tilde{\Phi} u$, $\abs{H}^{2} \hc{H} \Phi$, $\hc{(\cd_{\mu} \Phi)} (\cd^{\mu} H)$ \\
5 & $(\hc{\tilde{\Phi}} \ell)^{T} \mathcal{C} (\hc{\tilde{H}} \ell)$ \\
6 & $\abs{H}^{4} (\hc{H} \Phi)$, $\abs{H}^{2} \square (\hc{H} \Phi)$, $(\hc{H} \cd^{\mu} H)^{*} (\hc{H} \cd_{\mu} \Phi)$, $\abs{H}^{2} (\bar{\ell} \Phi e)$, $\abs{H}^{2} (\bar{q} \tilde{\Phi} u)$, $\abs{H}^{2} (\bar{q} \Phi d)$, $\hc{H} \Phi G_{\mu\nu}^{A} G^{A\mu\nu}$, $\hc{H} \Phi G_{\mu\nu}^{A} \tilde{G}^{A\mu\nu}$, $\hc{H} \Phi W_{\mu\nu}^{I} W^{I\mu\nu}$, $\hc{H} \Phi \tilde{W}_{\mu\nu}^{I} W^{I\mu\nu}$, $\hc{H} \Phi B_{\mu\nu} B^{\mu\nu}$, $\hc{H} \Phi \tilde{B}_{\mu\nu} B^{\mu\nu}$, $\hc{H} \tau^{I} \Phi W_{\mu\nu}^{I} B^{\mu\nu}$, $\hc{H} \tau^{I} \Phi \tilde{W}_{\mu\nu}^{I} B^{\mu\nu}$, $(\bar{\ell} \sigma^{\mu\nu} e) \tau^{I} \Phi W_{\mu\nu}^{I}$, $(\bar{\ell} \sigma^{\mu\nu} e) \Phi B_{\mu\nu}$, $(\bar{q} \sigma^{\mu\nu} T^{A} u) \tilde{\Phi} G_{\mu\nu}^{A}$, $(\bar{q} \sigma^{\mu\nu} u) \tau^{I} \tilde{\Phi} W_{\mu\nu}^{I}$, $(\bar{q} \sigma^{\mu\nu} u) \tilde{\Phi} B_{\mu\nu}$, $(\bar{q} \sigma^{\mu\nu} T^{A} d) \Phi G_{\mu\nu}^{A}$, $(\bar{q} \sigma^{\mu\nu} d) \tau^{I} \Phi W_{\mu\nu}^{I}$, $(\bar{q} \sigma^{\mu\nu} d) \Phi B_{\mu\nu}$,
$(\hc{H} i \overleftrightarrow{\cd}_{\mu} \Phi) (\bar{\ell} \gamma^{\mu} \ell)$, $(\hc{H} i \overleftrightarrow{\cd}_{\mu}^{I} \Phi) (\bar{\ell} \tau^{I} \gamma^{\mu} \ell)$, $(\hc{H} i \overleftrightarrow{\cd}_{\mu} \Phi) (\bar{e} \gamma^{\mu} e)$, $(\hc{H} i \overleftrightarrow{\cd}_{\mu} \Phi) (\bar{q} \gamma^{\mu} q)$, $(\hc{H} i \overleftrightarrow{\cd}_{\mu}^{I} \Phi) (\bar{q} \tau^{I} \gamma^{\mu} q)$, $(\hc{H} i \overleftrightarrow{\cd}_{\mu} \Phi) (\bar{u} \gamma^{\mu} u)$, $(\hc{H} i \overleftrightarrow{\cd}_{\mu} \Phi) (\bar{d} \gamma^{\mu} d)$, $(\hc{\tilde{H}} i \overleftrightarrow{\cd}_{\mu} \Phi) (\bar{u} \gamma^{\mu} d)$ 
\\
\multicolumn{2}{c}{Scalar Triplet $\vec{\Delta}$ DM decay ($Y_{\vec{\Delta}} = 0$). We denote $\Delta \equiv \Delta^{I} \tau^{I}$.} 
\\
3 & $\hc{H} \Delta H$ \\
4 & $\hc{\ell} \mathcal{C} \Delta \ell$, $\hc{q} \mathcal{C} \Delta q$ \\
5 & $\bar{\ell} \Delta H e$, $\bar{q} \Delta H d$, $\bar{q} \Delta \tilde{H} u$, $\abs{H}^{2} \hc{H} \Delta H$, $\hc{(\cd_{\mu} H)} \Delta (\cd_{\mu} H)$, $\bar{\ell} \Delta \slashed{\cd} \ell$, $\bar{q} \Delta \slashed{\cd} q$ \\
6 & $(\hc{\tilde{H}} \ell)^{T} \mathcal{C} (\hc{\tilde{H}} \Delta \ell)$, $\abs{H}^2 \hc{\ell} \mathcal{C} \Delta \ell$, $ \abs{H}^2 \hc{q} \mathcal{C} \Delta q$, $\hc{\ell} \mathcal{C} \sigma^{\mu\nu} \Delta \ell B_{\mu\nu}$, $\hc{\ell} \mathcal{C} \sigma^{\mu\nu} \ell \Delta^I W_{\mu\nu}^I$, $\hc{q} \mathcal{C} \sigma^{\mu\nu} \Delta q B_{\mu\nu}$, $\hc{q} \mathcal{C} \sigma^{\mu\nu} q \Delta^I W_{\mu\nu}^I$, $\hc{q} \mathcal{C} \sigma^{\mu\nu}  T^A \Delta q G_{\mu\nu}^A$, $\bar{\ell} \mathcal{C} \sigma^{\mu\nu} (\cd_\mu \cd_\nu \Delta) \ell$, $\bar{q} \mathcal{C} \sigma^{\mu\nu} (\cd_\mu \cd_\nu \Delta) q$, $(\bar{\ell} \gamma^\mu  e) (\Delta i \overleftrightarrow{\cd}_{\mu} H)$, $(\bar{q} \gamma^\mu d) (\Delta i \overleftrightarrow{\cd}_{\mu} H)$, $(\bar{q} \gamma^\mu u) (\Delta i \overleftrightarrow{\cd}_{\mu} \tilde{H})$ 
\\
\multicolumn{2}{c}{Scalar Triplet $\vec{\Delta}$ DM decay ($Y_{\vec{\Delta}} = 2$)}  
\\
3 & $\hc{H} \Delta \tilde{H}$ \\
4 & $\T{\ell} \mathcal{C} \Delta \epsilon \ell$ \\
5 & $\bar{\ell} \Delta \tilde{H} e$, $\bar{q} \Delta \tilde{H} d$, $\bar{q} \hc{\Delta} H u$, $\abs{H}^{2} \hc{H} \Delta \tilde{H}$, $\hc{(\cd_{\mu} H)} \Delta (\cd_{\mu} \tilde{H})$, $\bar{\ell} \Delta \slashed{\cd} \epsilon \ell^{*}$ \\
6 & $(\hc{\tilde{H}} \ell)^{T} \mathcal{C} (\hc{H} \Delta \ell)$, $\abs{H}^{2} \T{\ell} \mathcal{C} \Delta \epsilon \ell$, $\T{\ell} \mathcal{C} \sigma^{\mu\nu} \Delta \ell B_{\mu\nu}$, $\T{\ell} \mathcal{C} \sigma^{\mu\nu} \ell \Delta^I W_{\mu\nu}^I$,
$\T{\ell} \epsilon \sigma^{\mu\nu} (\cd_\mu \cd_\nu \Delta) \ell$,
$(\bar{\ell} \gamma^\mu e) (\Delta i \overleftrightarrow{\cd}_{\mu} \tilde{H})$, $(\bar{q} \gamma^\mu d) (\Delta i \overleftrightarrow{\cd}_{\mu} \tilde{H})$, $(\bar{q} \gamma^\mu  u) (\Delta i \overleftrightarrow{\cd}_{\mu} H)$
\\
\multicolumn{2}{c}{Fermion singlet $N$ DM decay (consistent with \cite{Haba:2010ag})} 
\\
4 & $\bar{\ell} \tilde{H} N$ \\
5 & -- \\
6 & $\abs{H}^{2} \bar{\ell} \tilde{H} N$, $(\bar{\ell} \sigma^{\mu\nu} N) \tau^{I} \tilde{H} W_{\mu\nu}^{I}$, $(\bar{\ell} \sigma^{\mu\nu} N) \tilde{H} B_{\mu\nu}$, $(\hc{H} i \overleftrightarrow{\cd}_{\mu} \tilde{H}) (\bar{e} \gamma^{\mu} N)$, $(\bar{e} \gamma_{\mu} N) (\bar{u} \gamma^{\mu} d)$, $(\T{\ell} \mathcal{C} \epsilon \gamma_{\mu} \tau^{I} \ell) (\bar{e} \gamma^{\mu} N)$, $(\bar{q} \gamma_{\mu} \epsilon q^*) (\bar{d} \gamma^{\mu} N)$, $(\bar{\ell} N) \T{(\bar{q} d)}$, $(\bar{\ell} N) (\bar{u} q)$ 
\\
\multicolumn{2}{c}{Fermion doublet $L$ DM decay} 
\\
4 & $\bar{L} H e$\\
5 & $(\hc{\tilde{H}} L)^{T} \mathcal{C} (\hc{\tilde{H}} \ell)$ \\
6 & $\abs{H}^{2} (\bar{L} H e)$, $(\bar{L} \sigma^{\mu\nu} e) H \tau^{I} W_{\mu\nu}^{I}$, 
$(\bar{L} \sigma^{\mu\nu} e) H B_{\mu\nu}$, $(\hc{H} i \overleftrightarrow{\cd}_{\mu} H) (\bar{L} \gamma^{\mu} \ell)$, $(\hc{H} i \overleftrightarrow{\cd}_{\mu}^{I} H) (\bar{L} \tau^{I} \gamma^{\mu} \ell)$, $(\bar{L} \gamma_{\mu} \ell) (\bar{\ell} \gamma^{\mu} \ell)$, $(\bar{L} \gamma_{\mu} \ell) (\bar{q} \gamma^{\mu} q)$, $(\bar{L} \gamma_{\mu} \tau^{I} \ell) (\bar{q} \gamma^{\mu} \tau^{I} q)$, $(\bar{L} \gamma_{\mu} \ell) (\bar{e} \gamma^{\mu} e)$, $(\bar{L} \gamma_{\mu} \ell) (\bar{u} \gamma^{\mu} u)$, $(\bar{L} \gamma_{\mu} \ell) (\bar{d} \gamma^{\mu} d)$, $(\bar{L} e) (\bar{d} q)$, $(\bar{L} e) \epsilon (\bar{q} u)$, $(\bar{L} \sigma_{\mu\nu} e) \epsilon  (\bar{q} \sigma^{\mu\nu} u)$ 
\\
\multicolumn{2}{c}{
Fermion triplet $\vec{\Sigma}$ DM decay ($Y_{\vec{\Sigma}} = 0$). We denote $\Sigma \equiv \Sigma^{I} \tau^{I}$.} 
\\
4 & $\bar{\ell} \Sigma \tilde{H}$ \\
5 & $\hc{H} \bar{e} \Sigma \tilde{H}$, $\bar{\ell} \gamma^\mu \Sigma \cd_\mu \tilde{H}$ \\
6 & $\abs{H}^{2} (\bar{\ell} \Sigma \tilde{H})$, $(\bar{\ell} \sigma^{\mu\nu} \Sigma^{I}) \tilde{H} W_{\mu\nu}^{I}$, $(\bar{\ell} \sigma^{\mu\nu} \Sigma) \tilde{H} B_{\mu\nu}$, $(\hc{H} i \overleftrightarrow{\cd}_{\mu} \tau^{I} \tilde{H}) (\bar{e} \gamma^{\mu} \Sigma^{I})$, $(\T{\ell} \mathcal{C} \epsilon \gamma_{\mu} \tau^{I} \ell) (\bar{e} \gamma^{\mu} \Sigma^{I})$, $(\bar{q} \gamma_{\mu} \tau^{I} \epsilon q^*) (\bar{d} \gamma^{\mu} \Sigma^{I})$, $(\bar{\ell} e)(\bar{\Sigma} \epsilon \ell^*)$, $(\bar{\ell} \Sigma) \epsilon \T{(\bar{q} d)}$, $(\bar{\ell} \Sigma) (\bar{u} q)$ 
\\
\multicolumn{2}{c}{
Fermion triplet $\vec{\Sigma}$ DM decay ($Y_{\vec{\Sigma}} = -2$) } 
\\
4 & $\bar{\ell} \Sigma H$ \\
5 & $\hc{H} \bar{\Sigma} e H$, $\T{\ell} \gamma^\mu \Sigma \cd_\mu \tilde{H}$ \\
6 & $\abs{H}^{2} (\bar{\ell} \Sigma H)$, $(\bar{\ell} \sigma^{\mu\nu} \Sigma^{I}) H W_{\mu\nu}^{I}$, $(\bar{\ell} \sigma^{\mu\nu} \Sigma) H B_{\mu\nu}$, $(\hc{H} i \overleftrightarrow{\cd}_{\mu} \tau^{I} H) (\bar{e} \gamma^{\mu} \Sigma^{I})$, $(\bar{\ell} \gamma_{\mu} \tau^{I} \ell) (\bar{e} \gamma^{\mu} \Sigma^{I})$, $(\bar{q} \gamma_{\mu} \tau^{I} q) (\bar{e} \gamma^{\mu} \Sigma^{I})$, $(\bar{q} \gamma_{\mu} \tau^{I} \epsilon q) (\bar{u} \gamma^{\mu} \Sigma^{I})$, $(\bar{\ell} e)(\bar{\Sigma} \ell)$, $(\bar{q} d) (\bar{\Sigma} \ell)$, $(\bar{q} u) \epsilon  \T{(\bar{\ell} \Sigma)}$ \\
\end{tabular}
\label{tab:dm:decay:ops}
\end{table}

Cosmology requires that the dark matter particles must have lifetime at least $10^{26}$~s. 
It is well known that lifetimes of that order may be achieved if the operators $d<6$ are forbidden \cite{Arvanitaki:2008hq}. 
In our framework we assume that non-Abelian flavor symmetries forbid the $d<6$ operators.   
The decay width of dark matter by $d=6$ operator is given by
\begin{equation}
\Gamma \sim \left(\frac{\mdm}{\Lambda}\right)^{4} \mdm I,
\end{equation}
and that by the $d=4$ operator by
\begin{equation}
\Gamma \sim \mdm I,
\end{equation}
where $\mdm$ is dark matter mass, $\Lambda$ is the scale of new physics and $I$ is a phase space integral.
If $\Lambda = \mg$ and $\mdm \sim \mathcal{O}(\text{TeV})$, then $d=6$ operators give a sufficiently long lifetime $\tau=10^{26}~\text{s}$ of the dark matter. If $d=6$ operators are suppressed by some smaller scale $\Lambda<\mg$, and/or $\mdm < 1~\text{TeV}$, then there must be additional suppression factors
to achieve long enough dark matter lifetime.
If $O_{d}$ is a $d=4$ operator, the needed suppression factor is $\epsilon^{n} \sim (1~\text{TeV}/10^{16}~ \text{GeV})^{4} \sim 10^{-52}$. 
We consider such small numerical coefficients in front of operators unnatural and disregard this possibility here.

\section{Breaking the flavor Group to Nothing}
\label{sec:breaking:to:nothing}

As a first step we aim to forbid $d < 6$ dark matter decay operators $O_\text{d}$ by judicious assignment of flavor multiplets to SM fields and dark matter. It is particularly simple in the case of fermion singlet dark matter: only one operator $\bar{\ell} \tilde{H} N$ has to be forbidden \cite{Haba:2010ag}.
Even so, there will be additional operators that arise from flavons $\phi_{i}$ that break the symmetry group. Thus there will be higher order terms of the form 
\begin{equation}
 O_{\text{d}} \phi_{i},\; O_{\text{d}} \phi_{i} \phi_{j},\; O_{\text{d}} \phi_{i} \phi_{j} \phi_{k},\; \ldots.
 \label{eq:flavon:dm:decay}
\end{equation}
When the flavons get VEVs and break the flavor symmetry, these terms generate the dark matter decay operators of the form
\begin{equation}
 \epsilon O_{\text{d}},\; \epsilon^{2} O_{\text{d}},\; \epsilon^{3} O_{\text{d}},\; \ldots.
\end{equation}
In most models, flavons are SM singlet scalars but transform nontrivially under the non-Abelian flavor group that they break near the cutoff scale. It is natural for the ratios $\epsilon_{i} \equiv \vev{\phi_{i}}/\Lambda$ to be $\order{1}$. To get mass hierarchy, usually a Froggatt-Nielsen flavon $\theta$ is added that transforms as the trivial singlet of the non-Abelian part of the flavor group but nontrivially under a $U(1)$ or $Z_{N}$ group. 
We know from phenomenology that parameter $\epsilon \equiv \vev{\theta}/\Lambda$ \emph{must} be larger than $\mathcal{O}(1/10)$, 
typically of the order of the Cabbibo angle 0.2.  This gives far too large a decay rate for dark matter.
For example, in the fermion singlet dark matter case, the effective operator $\bar{\ell} \tilde{H} N \phi$
 yields the $d = 4$ dark matter decay operator $\bar{\ell} \tilde{H} N$ with a coupling too large to allow the existence of dark matter.

How could one suppress the generated decay operators? The possible solutions are:
\begin{enumerate}
 \item If, in order to get an acceptable dark matter life time, one needs to forbid all terms \eqref{eq:flavon:dm:decay} to order $n$, one can choose a huge flavor group with rank $n+1$ and choose the SM and dark matter representations in the ``right" way such that the decay operators are suppressed by $\epsilon^n$ where
 $n$ is a large number. 
 However, doing that, all predictivity for fermion masses and mixing angles is likely to be lost because, typically, $n\sim 50$ is needed. 
 A principle of economy and simplicity
 can be used to argue against such a baroque group, although such a scenario can, in principle, work.
 
 \item The second possibility is that dark matter decay operators are suppressed by an additional factor of $1/\Lambda^{2}$ that arises from new physics beyond the 
 standard model but is not related to flavor physics. For example, dark matter can live in a hidden sector that communicates with the SM and flavons via heavy GUT (or Planck) scale messengers. Such a scenario is possible but is not related to new physics scenarios considered in this work.

 \item Thirdly, the r\^{o}le of flavons can be played by higher multiplets of SM like the Higgs doublet $H$  or the triplet Higgs $\Delta$ of type II seesaw \cite{magg:1980,lazarides:1981,mohapatra:1981,cheng:1980} that have VEVs at the \emph{electroweak} scale or lower.\footnote{Note that, in this case, the SM Higgs boson cannot be used in the Froggatt-Nielsen mechanism with new dynamics at the TeV scale \cite{Giudice:2008uua} because $\epsilon \sim \mathcal{O}(1/10)$. } Indeed, the SM Higgs boson VEV defines the electroweak scale. 
 Then $v/\Lambda \ll 1$ and the generated decay operators are suppressed. 
 \end{enumerate}
If the non-Abelian flavor group is completely broken, as we assume in this work, the last option is the only one that guarantees that the magnitude of dangerous  dark matter decay operators generated by the symmetry breaking remains phenomenologically acceptable. Luckily, this is also the only 
flavor physics scenario that can be directly testable at the LHC experiments. 

As we have emphasized several times before, in order to generate fermion mass hierarchies, any realistic flavor model must 
allow small and large flavor breaking VEVs to co-exist. We find two possibilities how this can happen.
\begin{enumerate}
 \item If the non-Abelian flavor group $G_\text{f}$ is broken in such a way that the full model does not have residual symmetries but
some well defined fermion sector of the model does have a remnant $Z_n$,
 there \emph{can} be flavons with GUT scale VEVs in one of these sectors. For example, suppose at least one $Z_2$ subgroup of $G_\text{f}$ remains unbroken in the neutrino sector. If $\text{DM} \sim -1$ under this $Z_{2}$, the large dark matter decay terms from neutrino sector flavons are forbidden. In the whole theory, $G_\text{f}$ is broken to nothing, so in this case the flavons in the charged lepton mass sector still must have electroweak scale VEVs.
 \item For a Froggatt-Nielsen flavon $\theta$ that transforms as a singlet of the non-Abelian part of the flavor group, it is possible to forbid the terms \eqref{eq:flavon:dm:decay} to all orders. If a dark matter decay operator $O_{\text{d}}$ is forbidden because it is a nontrivial singlet or a higher multiplet of $G_\text{f}$, then all terms $O_{\text{d}} \theta^n$ are still forbidden if $\theta$ is the trivial singlet of $G_\text{f}$. (If $\theta$ is a nontrivial singlet then $O_{\text{d}}$ must transform as a \emph{higher} multiplet.)
\end{enumerate}
 
 \section{An Example Model}
\label{sec:ex}
 
We will use the $T_{7}$ model \cite{Cao:2010mp} extended by a Froggatt-Nielsen flavon $\theta$ and dark matter to illustrate our findings.
The irreducible representations of $T_{7}$ are the trivial singlet $\mathbf{1_{0}}$, $\mathbf{1_{1}}$, $\mathbf{1_{2}}$, $\mathbf{3}$ and $\mathbf{\bar{3}}$. The tensor product of triplets is $\mathbf{3} \times \mathbf{3} = \mathbf{\bar{3}} + \mathbf{\bar{3}} + \mathbf{3}$. The Lagrangian must transform as the trivial singlet that is contained in the product $\mathbf{\bar{3}} \times \mathbf{3} = \mathbf{1_{0}} + \mathbf{1_{1}} + \mathbf{1_{2}} + \mathbf{3} + \mathbf{\bar{3}}$.

Consider a $T_{7}$ toy model where $\ell, H \sim \mathbf{3}$, $e_{i} \sim \mathbf{1_{i}}$ and the SM singlet fermion dark matter $N \sim \mathbf{1_{0}}$. To get charged lepton mass hierarchy, the $T_{7}$ group is extended to $T_{7} \times U(1)_{\text{FN}}$. The charges of right handed leptons under the $U(1)_{\text{FN}}$ for $e$, $\mu$ and $\tau$ are $2$, $1$ and $0$, respectively; the Froggatt-Nielsen flavon $\theta \sim \mathbf{1_{0}}$ has Froggatt-Nielsen charge $-1$. Below we will indicate $T_{7}$ representations by indices. Charged lepton Yukawa terms are given by
\begin{displaymath}
 \frac{\langle\theta_{\mathbf{1_{0}}}\rangle^{n_{i}}}{\Lambda^{n_{i}}} \bar{\ell}_{\mathbf{\bar{3}}} H_{\mathbf{3}} e_{\mathbf{1_{i}}},
\end{displaymath}
with $n_{i} = 2, 1, 0$. But the dark matter decay term
\begin{displaymath}
  \frac{\langle\phi_{\mathbf{1_{0}}}\rangle^{n}}{\Lambda^{n}} \bar{\ell}_{\mathbf{\bar{3}}} \tilde{H}_{\mathbf{\bar{3}}} N_{\mathbf{1_{0}}},
\end{displaymath}
is forbidden to all orders.

$T_{7}$ is a subgroup of $SU(3)$. If the latter is extended to $U(3)$, its $U(1)$ subgroup can serve as the Froggatt-Nielsen group.
Thus, in principle, in this scenario both the observed flavor physics and the existence of dark matter can originate form 
the underlying continuous non-Abelian symmetry.

\section{Discussion and Conclusions}
\label{sec:conclusions}

We have studied general requirements for models that simultaneously explain both the quark and lepton flavor structure and the stability of dark matter. 
Our work is based on the assumptions that the flavor group contains no extra $Z_{N}$ symmetries added by hand, that the full flavor
group is entirely broken to nothing, and that the flavor model explains both the mixing angles and mass hierarchies of the SM fermions.
We find that any flavor model with these properties  should contain both the higher multiplet flavons that generate the observed mixing patterns and the  flavor singlet Froggatt-Nielsen flavons. After the flavor symmetry breaking  the VEVs of the former must be at the electroweak scale, $v/\Lambda \ll 1,$ so that the dark matter particles have cosmologically acceptable lifetime, while the latter must have $\vev{\theta}/\Lambda$ of $\order{1/10}$ to generate the mass hierarchies.  This precludes the use of flavor doublet and triplet scalars in some kind of non-Abelian Froggatt-Nielsen mechanism \cite{Bazzocchi:2009pv} in our scenario of symmetryless dark matter.

To address the existence of dark matter, all dark matter decay operators of dimension $d<6$ must have been initially forbidden by the flavor group.
For the explained flavor symmetry breaking pattern the dark matter remains long-lived. Therefore the dark matter of the universe exists 
without any exact low energy symmetry.

Because the dark matter decay operators must transform as nontrivial singlets or higher multiplets under the flavor group, this approach favors larger non-Abelian groups with complex representations like $T_{7}$, $\Delta(27)$ or $\Delta(54)$ instead of smaller groups like $A_{4}$ or $S_{4}$ with additional factors of $Z_{N}$. The simplest way to embed dark matter in a flavor group is to consider fermion singlet dark matter: in this case only one dark matter decay operator has to be forbidden.

While it is hard to explain the large mixing angles in the lepton sector with the Froggatt-Nielsen mechanism, the latter is popular for generating the small Cabibbo angle of the quark sector. In this case quarks and the Froggatt-Nielsen flavon are charged under an Abelian $U(1)$ group. To forbid generation of too large dark matter decay operators, the $U(1)$ charge of dark matter would have to be very large, most probably not allowed by the requirement of perturbativity. Thus Froggatt-Nielsen models that give both quark mixing and stabilize dark matter are not favored.

To have the flavon VEVs at the electroweak scale or below, it is natural for the higher multiplet flavons to be higher representations of the SM gauge group as well: the SM Higgs doublets that break the electroweak symmetry or the triplet Higgs bosons of the type II seesaw mechanism can double as flavons. Unlike 
the singlet flavons with masses at the GUT or Planck scale, multiple SM Higgs bosons or triplet Higgses of type II seesaw can be discovered at the LHC, making the scenario of symmetryless dark matter directly testable.

\section*{Acknowledgments}
We thank Enrico Bertuzzo for discussions and the organizers of CERN TH-LPCC summer institute on LHC physics where part of this work was performed.
This work was supported by the ESF Mobilitas 2 grant MJD140, Mobilitas 3 grant MTT8, ESF grants 8090 and 8943, and SF0690030s09.

\bibliographystyle{h-physrev3} 
\bibliography{symmetryless_DM} 
\end{document}